\documentclass{mn2e}
\usepackage{graphicx}
\newcommand{\Fs}{\,^*\! F}
\newcommand{\bA}{\bmath{A}}

\newcommand{\bJ}{\bmath{J}}
\newcommand{\bj}{\bmath{j}}

\newcommand{\bv}{\bmath{v}}

\newcommand{\bB}{\bmath{B}}
\newcommand{\bE}{\bmath{E}}
\newcommand{\bH}{\bmath{H}}
\newcommand{\bD}{\bmath{D}}

\newcommand{\text}[1]{\quad\mbox{#1}\quad}

\newcommand{\vpr}[2]{\bmath{#1} \!\times\! \bmath{#2}}

\newcommand{\oder}[2]{\frac{d #1}{d #2}}

\newcommand{\Pd}[1]{\partial_{#1}}

\begin{document}
\title{Simulations of axisymmetric magnetospheres of  neutron stars}
\author[S.S.Komissarov]{S.S.Komissarov\\
Department of Applied Mathematics, the University of Leeds,
Leeds, LS2 9GT, UK.\\
e-mail: serguei@maths.leeds.ac.uk}
\date{Received/Accepted}
\maketitle

\begin{abstract}

In this paper we present the results of time-dependent simulations of dipolar
axisymmetric magnetospheres of neutron stars carried out both within the
framework of relativistic magnetohydrodynamics and within the framework of
resistive force-free electrodynamics.  The results of force-free simulations
reveal the inability of our numerical method to accommodate the equatorial
current sheets of pulsar magnetospheres and raise a question mark about the
robustness of this approach. On the other hand, the MHD approach allows to make
a significant progress.  We start with a nonrotating magnetically dominated
dipolar magnetospheres and follow its evolution as the stellar rotation is
switched on. We find that the time-dependent solution gradually approaches the
steady state that is very close to the stationary solution of the Pulsar
Equation found by Contopoulos et al.(1999). This result suggests that other
stationary solutions that have the y-point located well inside the light
cylinder are unstable.  The role of the particle inertia and pressure on the
structure and dynamics of MHD magnetospheres is studied in details, as well as
the potential implications of the dissipative processes in the equatorial
current sheet. We argue that pulsars may have differentially rotating
magnetospheres which develop noticeable structural oscillations and that this
may help to explain the nature of the sub-pulse phenomena.
    
\end{abstract}
\begin{keywords}
stars:pulsars:general -- stars:winds and outflows -- 
relativity -- MHD
\end{keywords}

\section{Introduction} 

Pulsar magnetospheres are complicated electrodynamic systems that involve the
acceleration of charged particles by the rotationally induced electric field,
the electron-positron pair production, and the bulk outflow of magnetospheric
plasma through the light cylinder.  The pair production presumably occurs on the
scale which is significantly smaller compared to the magnetospheric scale (the
last is determined by the light cylinder radius $\varpi_{lc}=c/\Omega$). This
suggests that that the global structure of pulsar magnetospheres may not depend
on the details of the pair production mechanism which may enter the problem only
in the form of boundary conditions.  The simplest assumption is that the plasma
supply is sufficiently plentiful to allow the MHD description throughout the
whole magnetosphere. Although this may be not so in the case of slowly rotating
or weakly magnetized pulsars \cite{HiA}, it still makes sense to regard the MHD
solution as a zero approximation model and a good reference point.
                                                                                             
Although pulsar magnetospheres are generically three dimensional it is still
reasonable to start with a much simpler axisymmetric case as it involves most of
the basic physical elements of the general case.  Michel\shortcite{Mic73} and
Scharlemann \& Wagoner~\shortcite{ScW} first analysed the structure of the
stationary axisymmetric pulsar magnetosphere in the limit of force-free
electrodynamics where both the pressure and inertia of magnetospheric plasma are
assumed to be vanishingly small. They found that in this case the problem is
reduced to the following second order PDE that is now called the Pulsar
Equation,  
\begin{equation}
(1-x^2)\left( 
    \frac{\partial^2 \Psi}{\partial x^2 }
    +\frac{1}{x}\frac{\partial \Psi}{\partial x}
    +\frac{\partial^2 \Psi}{\partial y^2}
    \right)
    -\frac{2}{x}\frac{\partial \Psi}{\partial x} = 
    -{\cal A}(\Psi)\oder{\cal A}{\Psi},
\end{equation}
where $x=\varpi/\varpi_{lc}$ and $y=z/\varpi_{lc}$ are the cylindrical
coordinates normalised to the radius of the light cylinder and $\Psi$ is the
magnetic flux function. This equation also involves the poloidal current
function, ${\cal A}(\Psi)$, that is unknown. If this function is somehow
specified the pulsar equation become a linear elliptic equation.  However, the
question of what actually determines the electric current function is not that
obvious and is still debated \cite{Mes,Bes05}.  The physical meaning of these
functions is quite simple: $\Psi =\Phi/2\pi$, where $\Phi(x,y)$ is the magnetic
flux through the axisymmetric circular loop given by $x=$const, $y=$const;
${\cal A}=2I/c$ where $I(x,y)$ is the total electric current flowing through the
same loop. It is also worth mentioning that if we set $A_\phi(x=0)=0$ then
$\Psi=A_\phi$ and that in a steady-state ${\cal A}=H_\phi$, where $A_\phi$ and
$H_\phi$ are the covariant azimuthal components of the vector-potential $\bA$
and the magnetic field $\bH$ in the non-normalised basis of cylindrical or
spherical coordinates.
 
Another important feature of this equation is the mathematical singularity at
the light cylinder. It was argued by Ingraham\shortcite{Ing} that the condition
of smooth passage through this surface together with the appropriate boundary
conditions determine the unique electric current function of the pulsar equation
(This property makes the problem somewhat similar to the classical eigenvalue
problem in the theory of differential equations.)  Ingraham\shortcite{Ing} even
proposed an iterative algorithm for finding this function. In the same year
Michel\shortcite{Mic73} did actually find an exact solution to the pulsar
equation in the case of split-monopole magnetic field that passed through the
singular surface both continuously and smoothly.  Although Michel did not use
Ingraham's approach his result supported Ingraham's idea and raised expectations
for finding smooth continuous solutions in more complicated realistic cases,
like for example the magnetosphere of a rotating dipole, as well.  (The recent
observations of magnetospheric eclipses in the binary pulsar J0737-3039 have
confirmed that the dipolar model may indeed be quite accurate \cite{LyT}.)

The electric current of Michel's solution has the same sign within the
magnetosphere with the exception of the equatorial current sheet that provides
current closure. Further analysis by Ingraham\shortcite{Ing} and
Michel\shortcite{Mic74} suggested that if the force-free solution for an
axisymmetric rotating dipole existed then far beyond the light cylinder it
should still closely resemble the split-monopole one. However, in the near zone
the structure of the dipole magnetosphere was expected to be qualitatively
different. Indeed, close to the star the effects of its rotation are relatively
small and so must be the difference between the solutions for the rotating and
the non-rotating magnetospheres. As the result, the so-called ``dead zone''
should exist where the magnetic field lines remain closed and the magnetospheric
plasma co-rotates with the star.

However, the problem of finding the self-consistent global solution for dipolar
magnetospheres turned out to be rather involved, even for the axisymmetric case
of aligned rotator. For a very long time, the only attempts to construct such
solutions involved utilisation of prescribed electric current functions.  Such
solutions would either exhibit kinks at the light cylinder or violate the
conditions of force-free approximation at some relatively short distance from
it, e.g. \cite{Mic82,BGI}.  The break through came only recently then
Contopoulos et al. (1999, hereafter CKF) have finally managed to solve the problem
numerically. In this solution the dead zone continues all the way to the light
cylinder where the so-called ``Y-point'' appears in the magnetic field
structure whereas beyond the light cylinder it has the same topology as the 
split-monopole solution including an infinitely thin equatorial current sheet. 
The return current of the equatorial current sheet splits 
at the Y-point into two currents sheets that flow along the surface separating
the dead zone from the open field zone. An additional finite width layer of
return current exists just outside of the dead zone and around the equatorial plane.

Uzdensky \shortcite{Uzd03} carried out asymptotic analysis of the force-free solutions 
in vicinity of the Y-point. It has turned out that in the presents of separating 
current sheets, similar to those  found in \cite{CKF}, the electromagnetic field 
becomes infinitely strong at the Y-point. He argued that this is unphysical and 
questioned correctness of the CKF-solution. As an alternative Uzdensky \shortcite{Uzd03}
considered the case with no current sheets, that is the case where the electric 
current returns within the equatorial layer of the open field zone only. 
In this case the electromagnetic field does not diverge at the Y-point but just 
outside of this point the electric field becomes stronger than the magnetic field
signaling a breakdown of the force-free approximation. No such complications 
were found for the Y-point located inside the light cylinder. In fact, in this 
case the separating surface crosses the equator at the right angle so that 
the Y-point turns into a "T-point".     

Gruzinov\shortcite{Gru05} also analysed the force-free solution in the vicinity of the 
Y-point located at the light cylinder in the presence of the current sheets 
and confirmed that in this case the electromagnetic field diverges at the Y-point.  
His analytical solution exhibits the angle of $77^o.3$ between the equatorial plane 
and the separating surface at the Y-point.  In order to verify this he repeated 
the CKF-calculations with much higher spatial resolution and the results seemed 
to confirm the development of a singularity with the expected inclination angle of
separatrices.      

The next potentially very important step was made by Goodwin et
al.\shortcite{Goo} who realised that the dead zone does not have to extend all
the way to the light cylinder but can be much smaller.  Such solutions select
their own electric current functions, smaller dead zones corresponding to weaker
currents.  In fact, the first solutions of the pulsar equation which described
both the equatorial current sheet and the dead zones that were located deep
inside the light cylinder were found by Lyubarskii\shortcite{Lyu90}.  However,
these solutions utilized a prescribed electric current function and, as the
result, exhibited the breakdown of the force-free approximation -- at a certain
distance from the light cylinder the electric field became stronger than the
magnetic field.   Goodwin et al.\shortcite{Goo} also included finite gas 
pressure inside the dead zone and showed that this allows solutions that 
remain nonsingular at the Y-point even when this point is located on the 
light cylinder. 
In this case their solution gives the inclination angle of the Y-point of $56^o.5$.
However, the inertia associated with the gas pressure has not been taken into 
account and thus the self-consistency of such approach is questionable.        

This idea of  Goodwin et al.\shortcite{Goo} was further explored by 
Timokhin\shortcite{Tim} who has constructed numerical solutions for 
force-free magnetospheres with the Y-point located within the light cylinder 
(with no finite gas pressure in the dead zone).  
He also pointed out that the pulsar spin-down rate depended not only on the size the
light cylinder but also on the size of the dead zone and if the ratio of these
scales were to evolve with the pulsar age then this could explain why the 
observed braking index of pulsars is smaller than the one derived from the 
models with the dead zone extending all the way up to the light cylinder.

Another interesting twist has been added by Contopoulos\shortcite{Con05}, who
argued that during the pulsar evolution the potential gap separating the star
surface from the polar cap magnetosphere grows in magnitude leading to a slower
rotation of the open field lines compared to the star and the dead zone. In such
a case, there is no a single light cylinder for the whole magnetosphere.
Instead, the dead zone has its own light cylinder that is located inside the
light cylinder of the polar cap.  To illustrate this point Contopoulos(2005)
constructed a set of numerical models of pulsar magnetospheres of this kind.
The evolution of the angular velocity of the polar cap magnetosphere with pulsar
age may also result in the lower values of the pulsar's braking index.

In spite of this remarkable progress the method of pulsar equation has its
obvious limitations. It does not address the question of stability of the
steady-state solutions and it does not allow to study possible non-stationary
phenomena in pulsar magnetospheres. Moreover, the numerical techniques that have
been used to solve the pulsar equation may turn up to be rather difficult to
adapt to the fully 3-dimensional problem of the oblique rotator. The obvious way
of approaching these problem is to relax the stationarity condition and to solve
the original system of time-dependent equations. Although the force-free
electrodynamics was used to model various astrophysical systems since 1970s the
focus was entirely on the steady-state equations. Only recently, the
time-dependent equations have been subjected to a systematic study
\cite{Uch,Gru99,Kom02a,Pun}. As the result it has been found that they form a
simple hyperbolic system of conservations laws in many respects similar to
relativistic MHD but only with the fast and the Alfv\'en hyperbolic waves
\cite{Kom02a}. Thus, a variety of standard numerical methods can be used to deal
with these equations. The very first numerical simulations of this kind, of the
monopole magnetospheres of black holes, seemed to confirm the suitability of
this approach \cite{Kom01}.  However, further applications to somewhat more
complex magnetic configurations revealed its limitations as the force-free
approximation would often break down following development of strong current
sheets \cite{Kom02a,Kom02b,Spi04,Asa}.  These results suggested to look for a
more general mathematical framework that would allow to handle current sheets
via introducing new channels of back reaction of plasma on the electromagnetic
field.  One of the most suitable options is the framework of resistive
relativistic MHD. However, this approach remains rather poorly studied even now.
Ideal relativistic MHD is a somewhat easier option and thanks to the recent
effort by several groups, a significant expertise has been obtained in
developing numerical methods for this system.  This approach allows to take into
account both the thermodynamic pressure of plasma heated in the current sheets
but the dissipation of electromagnetic energy is entirely due to numerical
resistivity that is not fully satisfactory.

Another alternative is resistive force-free electrodynamics with physical or
artificial resistivity. If the current sheets of pulsar magnetospheres are
indeed dissipationless then this approach might work provided that the utilized
Ohm law allows evolution towards a dissipationless force-free state.  The recent
time-dependent simulations by A.Spitkovsky show that this approach may indeed be
quite productive (the results were presented last summer at the conference on
``Physics of Astrophysical Outflows and Accretion Disks'', KITP, Santa-Barbara,
2005.)

In this paper we report the results of new time-dependent axisymmetric
simulations of rotating dipole magnetospheres of neutron stars within the
frameworks of resistive force-free electrodynamics and ideal relativistic
MHD. We only consider the case where the whole magnetosphere rotates with the
same angular velocity and thus our results are not relevant to the model of
Contopoulos\shortcite{Con05}. This model will be a subject of a separate study.
Note, that throughout the paper the graphic data are shown in the dimensionless
units where the magnetic dipole moment $\mu=1$, the angular velocity of the star
$\Omega$, and the speed of light $c=1$.  Hence the cylindrical radius of the
light cylinder $\varpi_{lc} =1$.

\section{Common features of the simulations} 

Using the standard notation of the 3+1 approach the metric form of a general
space-time can be written as

\begin{equation}
  ds^2 = (\beta^2-\alpha^2) dt^2 + 2 \beta_i dx^i dt + \gamma_{ij}dx^i dx^j,
\label{metric}
\end{equation}
where $\gamma_{ij}$ is the metric tensor of ``the absolute space'', $\alpha$ is
the lapse function, and $\bbeta$ is the shift vector.  For most purposes in
physics of pulsar magnetospheres the flat space approximation suffices and one
can enjoy the benefits of a global inertial frame where $\alpha=c$ and
$\beta_i=0$ (see however Beskin 1990 and Muslimov \& Tsygan 1990).  This is
exactly what was adopted in the numerical models of pulsar magnetospheres
described in the Introduction and in this study as well. However, the computer
codes used in the simulations are designed to work with a rather general
axisymmetric and stationary space-times. Moreover, in the case of oblique
rotator the frame rotating with the star seem to be more suitable as the
solution may become stationary in this frame. In such a case $\bbeta$ is no
longer vanishing and in the basis of spherical spacial coordinates
$$ \beta_r=\beta_\theta=0,\ \beta_\phi=c\Omega\sin^2\!\theta r^2,
$$ and hence
$$ \alpha^2=c^2, \text{and} \beta^2=c^2\Omega^2\sin^2\!\theta r^2,\
$$ where $\Omega$ is the angular velocity of the frame.

In our simulations, the computational grid covers the axisymmetric domain
$(r,\theta)=[r_{in},r_{out}]\times[0,\pi]$ so no symmetry condition is enforced
at the equatorial plane. The cell size in the $r$-direction is such that the
corresponding physical lengths in both directions are equal, $\Delta r = r
\Delta\theta$.  The ``radiative'' outer boundary, $r=r_{out}$, is always located
so far away from the star that the light signal does not cross the computational
domain by the end of the simulations -- this ensures that waves produced near
the star do not get reflected of the outer boundary at any rate.
 
As the Courant stability condition requires $\Delta t < \Delta r/c$, the
suitable time step for the outer part of the computational domain is much larger
than the one for its inner part. This allows us to reduce the computational cost
of simulations via splitting the computational domain into a set of rings such
that the outer radius of each ring is twice its inner radius and advance the
solution for each $k$th ring with it its own time step, $\Delta t_k$, such that
$\Delta t_{k+1} = 2\Delta t_k$. Thus, one integration step of ring $k$
corresponds to two integration steps of ring $k-1$, four integration steps of
ring $k-2$ and so on. As the result, the outer regions of the computational
domain are progressively less expensive in terms of CPU time.  This approach has
already been successfully applied in recent MHD simulations of Pulsar Wind
Nebulae \cite{KoL}.

The initial solution describes a nonrotating magnetosphere with exactly dipolar
magnetic field,
\begin{eqnarray} 
\nonumber B^{\hat{\phi}} &=& 0; \\
\label{dipole}
B^{\hat{r}} &=& 2 \mu \cos\theta/r^3;\\ \nonumber B^{\hat{\theta}} &=& \mu
\sin\theta/r^3,
\end{eqnarray} 
where $\mu$ is the magnetic dipole moment. At time $t=0$ the stellar rotation is
suddenly switched on via an appropriate inner boundary condition. The
simulations are then proceeded till it becomes clear whether the time-dependent
solution relaxes to a steady state on scales comparable with the light cylinder
radius or not.  Steady-state solutions that are found in such a way are
automatically stable to axisymmetric perturbations with wave-lengths exceeding
the cell size.

\section{Electrodynamic model} 

\subsection{Equations}

Following Komissarov\shortcite{Kom04a} we write vacuum Maxwell's equations as

\begin{equation}
  \Pd{i}(\sqrt{\gamma} B^i) =0 ,
\label{divB}
\end{equation}

\begin{equation}
(1/c)\Pd{t}B^i+e^{ijk}\Pd{j}E_k =0,
\label{Faraday}
\end{equation}

\begin{equation}
\label{divD}
  \Pd{i}(\sqrt{\gamma} D^i) = 4\pi\sqrt{\gamma} \kappa,
\end{equation}

\begin{equation}
   -(1/c)\Pd{t}D^i+e^{ijk}\Pd{j}H_k =(4\pi/c) J^i.
\label{Ampere}
\end{equation}
Here $\gamma =\det(\gamma_{ij})$ is the determinant of the metric tensor of the
absolute space, $e_{ijk} = \sqrt{\gamma} \epsilon_{ijk} $ is the Levi-Civita
tensor of the absolute space ($\epsilon_{123}=1$ for right-handed systems and
$\epsilon_{123}=-1$ for left-handed ones.)  The electric field, $\bE$, and the
magnetic field, $\bB$, are defined via
\begin{equation}
  E_i=\frac{\alpha}{2} e_{ijk}\Fs^{jk} ,
\end{equation}
and
\begin{equation}
  B^i=\alpha \Fs^{it},
\end{equation}
where $\Fs^{\mu\nu}$ is the Faraday tensor of the electromagnetic field, which
is simply dual to the Maxwell tensor $F^{\mu\nu}$:
                                                                                
\begin{equation}
\Fs^{\alpha \beta} = \frac{1}{2} e^{\alpha \beta \mu \nu} F_{\mu \nu}
\label{dual_F}
\end{equation}
where
                                                                                
\begin{equation}
e_{\alpha \beta \mu \nu} = \sqrt{-g}\,\epsilon_{\alpha \beta \mu \nu},
\label{LCt}
\end{equation}
is the Levi-Civita alternating tensor of spacetime.

\begin{figure}
\fbox{\includegraphics[width=80mm]{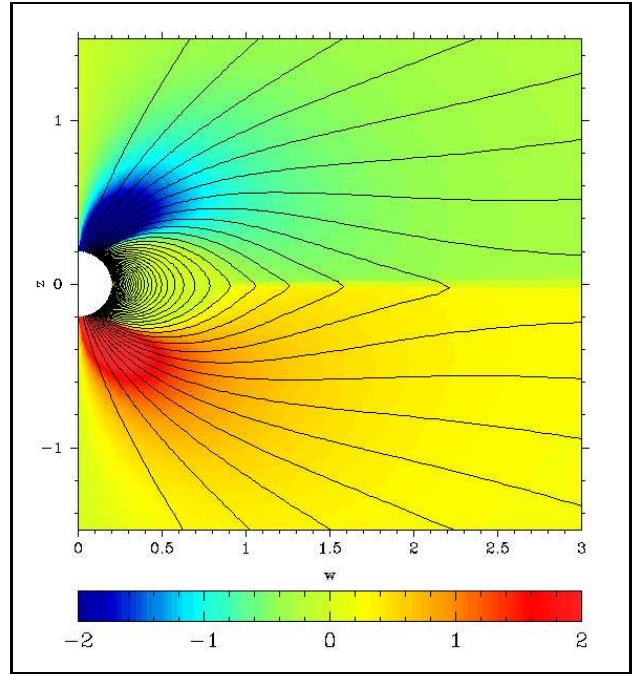}}
\caption{ Representative force-free solution. The contours 
show the magnetic flux surfaces and the colour image shows $H_\phi$.  
Notice that the magnetic field lines are closed even beyond the light 
cylinder, $\varpi_{ls}=1$.}
\label{ffde}
\end{figure}

The differential equations (\ref{divB}-\ref{Ampere}) are supplemented with the
following constitutive equations \cite{Kom04a}:
                                                                         
\begin{equation}
    c E_k = \alpha D_k + e_{kij}\beta^i B^j,
\label{E3}
\end{equation}

\begin{equation}
    c H_k = \alpha B_k - e_{kij}\beta^i D^j.
\label{H3}
\end{equation}

\begin{equation}
     J^k=(\alpha/c) j^k - \kappa \beta^k.
\label{e5}
\end{equation}
Note, that 1) the components of all vectors and tensors appearing in these
equations are measured in the non-normalised coordinate basis, $\{\Pd{i}\}$, of
spacial coordinates; 2) $\bD$, $\bB$, $\bj$, and $\kappa$ are the electric
field, the magnetic field, the electric current density, and the electric charge
density as measured by a local observer at rest in the absolute space. The
4-velocity of this observer is
$$ n^\nu = \frac{1}{\alpha}(c,\beta^i) ;
$$ In the inertial frames of flat spacetime $\bE=\bD$ and $\bB=\bH$.

The final equation that is needed to close the system is the Ohm law.  In
strongly magnetized plasma the conductivity is no longer isotropic and under
rather general conditions

\begin{equation}
  \bj = \sigma_\parallel \bD_\parallel + \sigma_\perp \bD_\perp + \bj_d,
\label{Ohm}
\end{equation}
where
\begin{equation}
    \bj_d = \kappa c \frac{\vpr{D}{B}}{B^2}
\label{jdrift}
\end{equation}
is the drift current.  Unless the magnetosphere has scarce supply of
electrically charged particles ,``charge starved'', the parallel conductivity is
very larger that leads to small residual parallel component of electric field,
\begin{equation}
\sigma_\parallel \rightarrow \infty, \text{ } \bD_\parallel \rightarrow 0.
\end{equation}
However, the strong magnetic field of pulsar magnetospheres effectively
suppresses conductivity across the magnetic field lines (unless the electric
field is even stronger than the magnetic one; this may be the case inside some
current sheets.)  So we can put

\begin{equation}
   \sigma_\perp=0 \text{if} B^2>D^2.
\end{equation}
As shown in \cite{Kom04a}, in this limit we approach the approximation of 
force-free electrodynamics. 

Within current sheets this simple prescription is unlikely to hold. 
On one hand, in the area of high current density one may expect strong 
anomalous resistivity leading to significantly reduced $\sigma_\parallel$.
As the result, the current sheet may become unstable to the tearing mode of 
reconnection \cite{Lyt03}. Since our intention at this point is merely to see 
if we can construct idealized  numerical solutions that are more-or-less 
force-free and thus can be compared with the steady state solutions 
of the Pulsar Equation we will ignore this physical effect for the time being.  
On the other hand, the mere symmetry of the aligned rotator magnetosphere suggests
that sooner or later the magnetic field of force-free solution will become 
very small in the equatorial plane beyond the light cylinder, thus leading to   
unavoidable breakdown of the $D^2<B^2$ condition of the force-free approximation.    
In such a case of relatively weak magnetic field the conductivity is expected to 
become much less anisotropic and we will assume that 

\begin{equation}
   \sigma_\perp=\sigma_\parallel \gg 1 \text{if} B^2<D^2.
\end{equation}
Notice, that the expression (\ref{jdrift}) for the drift current should also be
modified in this case as it implies that the drift speed
\begin{equation}
    \bv_d = c \frac{\vpr{D}{B}}{B^2}
\label{vdrift}
\end{equation}
becomes higher then the speed of light. We have tried several modifications for
the $v_d$ including
\begin{equation}
    v_d = c \frac{\vpr{D}{B}}{\max(B^2,D^2)}
\end{equation}
which is likely to underestimate $v_d$ in the current sheet and
\begin{equation}
    v_d = c \frac{\vpr{D}{B}}{|\vpr{D}{B}|} \text{if} B^2<D^2
\label{vdrift1}
\end{equation}
which certainly overestimates it. The actual value of conductivity in the current 
current sheet remain a free parameter. 

To find numerical solutions of Maxwell's equation, in particular the solutions
that are presented in the next section, we used the Godunov-type numerical
scheme described in Komissarov\shortcite{Kom04a}.
 
\subsection{Numerical simulations}

The most striking and at first somewhat perplexing result of our electrodynamic
simulations of pulsar magnetospheres is demonstrated in figure~\ref{ffde} -- contrary 
to what is found in the steady-state solutions of the pulsar equation the
magnetic field lines remain close even outside of the light cylinder. 
Moreover, the solution seems to be rather insensitive to the details of the model for
$\sigma_\perp$ and remains qualitatively the same even in the case $\sigma_\perp
=0$ (In fact, the solutions exhibited convergence in the limit $\sigma_\perp \to
0$.)  

The question of whether the field lines extending beyond the light cylinder should 
open up or not is in fact rather involved. One of the arguments often put forward in its 
discussion concerns the requirement for the speed of charged particles that fill the 
magnetosphere to remain smaller than the speed of light.    
In the drift approximation the charged particles move along the rotating magnetic 
field lines like beads on wire -- their motion is a composition of 
the rotational motion of the field line (the ``wire'') and the sliding motion of 
the particle (the ``bead'') along the field line. Within the light cylinder the speed of 
the rotational motion is less than the speed of light and the bead does not have to 
slide along the wire, but this is no longer the case beyond the light cylinder.  
Here the total speed of the bead can remain less than $c$ only if it also slides 
along the wire and only if the wire is twisted in the azimuthal direction (In such
a case the sliding motion may help to reduce the azimuthal component of the bead velocity.)   
These conditions may well be satisfied everywhere along the open field lines but not 
along the closed ones.  Indeed, such field lines cannot have an azimuthal component 
in the equatorial plane because of the symmetries of the problem. 
Thus, one may expect spinning up of the charged particles till their inertia becomes 
important and the centrifugal force opens up the closed field lines. 

On the other hand, the drift approximation itself may breakdown 
and the charges may start moving across the magnetic field well before their inertia 
becomes dynamical significant. One can look at this breakdown 
of the drift approximation from another prospective. Beyond the light cylinder 
the electric field of a force-free solution becomes stronger than the poloidal 
component of the magnetic field and thus stronger than the total magnetic field 
in regions where the azimuthal component of magnetic field vanishes. Such a strong 
electric field is capable of "tearing" charged particle off the magnetic field lines 
and driving strong electric current across the magnetic field. This is a micro-physical 
argument but there is an equally compelling macroscopic one.  Full opening of 
magnetic field lines implies an infinitely thin equatorial current sheet with an infinitely 
high electric current density. Even very small but finite resistivity will destroy this 
ideal configuration and result in a steady-state current layer of finite thickness and 
closure of some of the magnetic field lines in this layer. Thus, the only meaningful 
question is how many field lines will be closed in this layer. 

In order to explain the remarkable indifference of our solutions to the the value 
of $\sigma_\perp$ it is instructive to consider a much simpler
problem of a one dimensional current sheet.  Here we adopt an inertial frame
with Cartesian coordinates so that ($\bD=\bE$ and $\bH=\bB$).  Assuming that
$\bB=(0,B^y,0)$ and $\bE=(0,0,E^z)$ and using the following model for the
cross-field conductivity
\begin{equation}
   \sigma_\perp = \left\{
    \begin{array}{lcr}
        \sigma_0 &\text{if}& E^2>B^2\\ 0 &\text{if}& E^2 \leq B^2
    \end{array} \right. 
\label{sigma-perp}
\end{equation}
we find the following solution
$$ B^y = \left\{
    \begin{array}{lcr}
        -B &\text{if}& x < -1/\sigma_0\\ B\sigma_0 x &\text{if}& -1/\sigma_0 < x
         < 1/\sigma_0\\ B &\text{if}& x > 1/\sigma_0
    \end{array} \right. ,
$$
$$ 
E^z=B.
$$ 
This solution exhibits a number of interesting features.  First of all it is
stationary. Secondly, the magnitude of $\sigma_0$ effects only the width of the
current sheet. This may explain the observed convergence of the magnetospheric
solutions in the limit $\sigma_\perp\rightarrow 0$.  Finally, the
electromagnetic energy flows into the current sheet with the speed of light and
disappears inside of it. This property clearly exposes the nature of Ohm's law
(\ref{Ohm}) with prescription (\ref{sigma-perp}) or similar and outlines the
limitations of its applicability.  Indeed, in real plasma the Ohmic dissipation
would cause strong heating and increase of the gas pressure in the current
sheet.  This pressure would slow down the inflow of plasma into the current sheet
and significantly modify its structure.  However, this factor, as well as the
plasma inertia, are completely ignored in our version of Ohm's law. 

The fact that our numerical magnetospheric solution remains qualitatively the
same even in the limit $\sigma_\perp =0$ suggests that the numerical resistivity
of our scheme takes over in this limit. Different numerical schemes have
different dissipative properties and this may explain why the recent electrodynamic
simulations of Spitkovsky (which have been presented at various astrophysical
meetings) exhibit opening up of the field lines beyond the light cylinder and 
development of the dissipationless equatorial current sheet.  

The properties of the current sheets of real pulsars are determined by a number of 
competing micro-physical processes that might be rather difficult to account 
for in a macroscopic model. Our experiments with the simplistic generalised Ohm law 
(\ref{Ohm}--\ref{vdrift1}) show that it facilitates development of strongly dissipative 
current sheets. Since the dissipated energy of the electromagnetic field is not stored 
in any dynamical component and simply vanishes from the system this model could only  
be relevant for magnetospheres with effective radiative cooling. The strong 
gamma-ray emission from some pulsars may be interpreted as an indication of 
strong radiative current sheets  (see the Discussion). 
Unfortunately, such emission is not a common feature of pulsars.

\begin{figure*}
\fbox{\includegraphics[width=85mm]{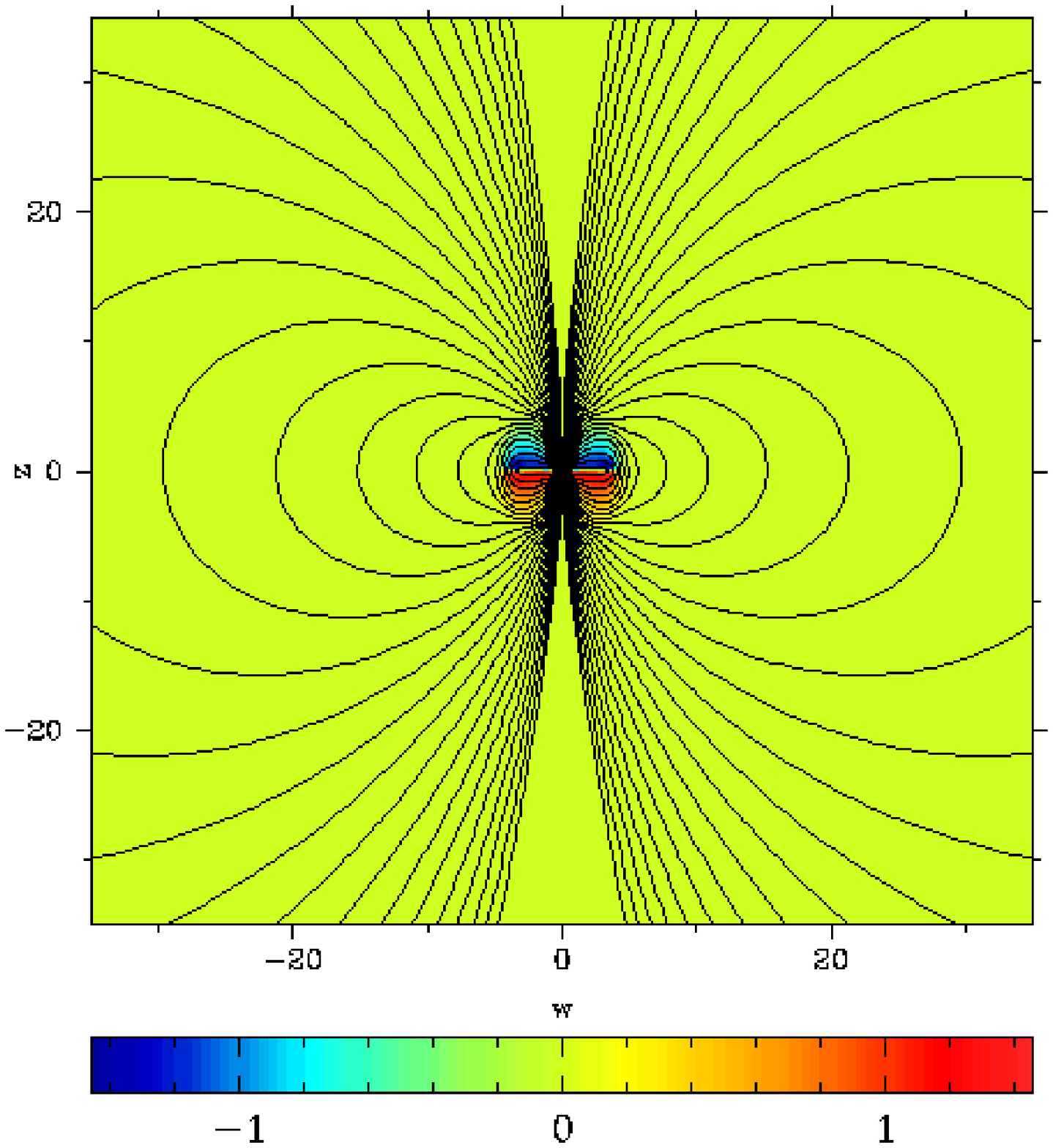}}
\fbox{\includegraphics[width=85mm]{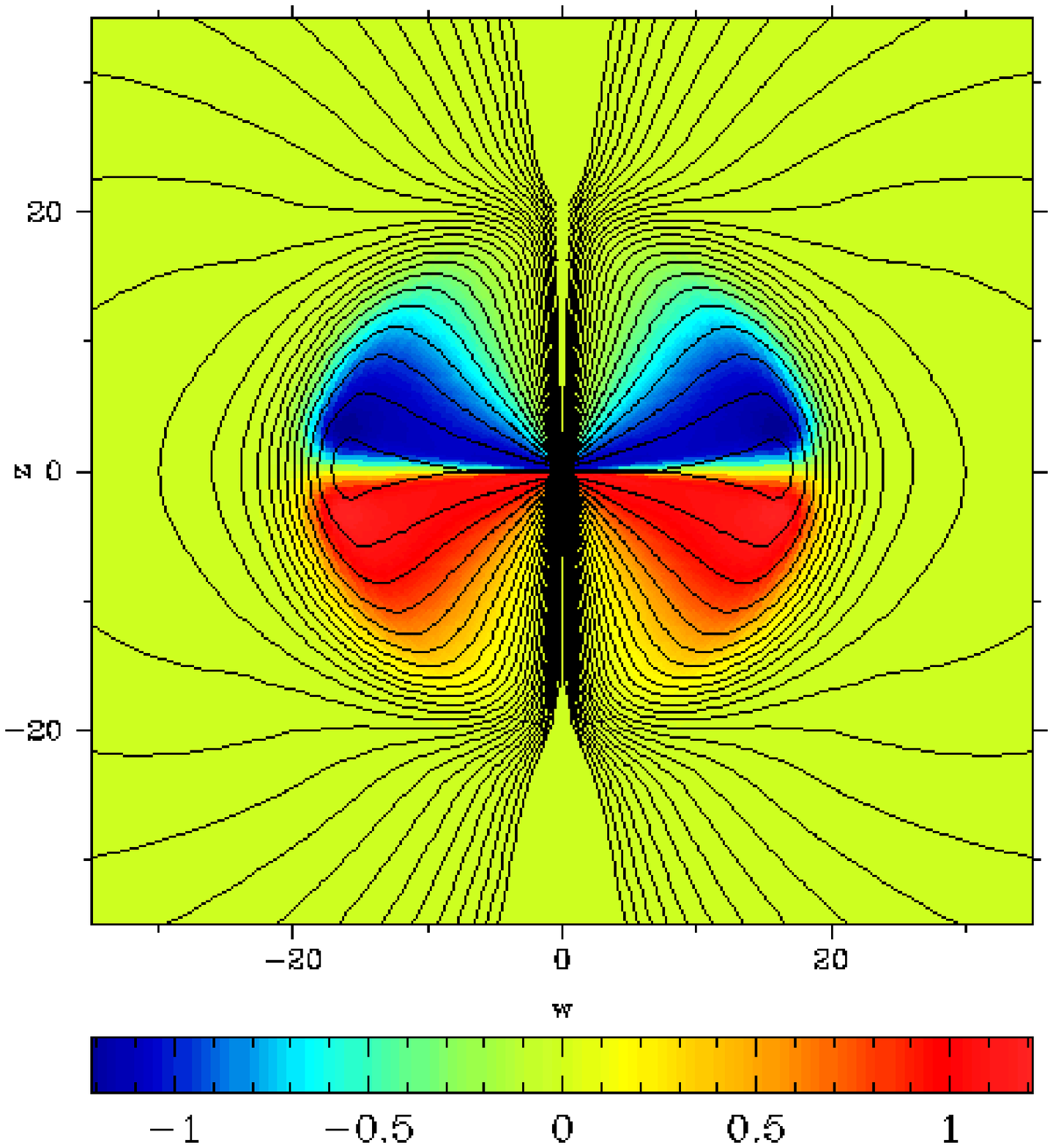}}
\caption{The evolution of spinned up dipole. This figure shows the solution at
$t=5$ (left panel) and at $t=20$ (right panel). The contours show the magnetic
flux function, $log_{10}\Psi$ and the colour image shows $H_\phi=2(I+I_d)/c$,
where $I$ is the electric current and $I_d$ is the displacement current.}
\label{glob}
\end{figure*}

\section{MHD model} 

The difficulties that we have encountered in dealing with current sheets in
force-free electrodynamics have forced us to return to the framework of full
relativistic MHD even if the pulsar magnetospheres are expected to be
magnetically dominated everywhere else and in spite of the fact that the system
of MHD equations becomes stiff in this regime. The MHD approximation takes into
account not only the gas pressure but also its inertia which may become
important both far away from the star, in the wind zone, and near the light
cylinder.  Ideally we should also incorporate a physical model for electric
resistivity which would allow us to conduct a proper study of the dissipation
within the equatorial current sheet but it makes perfect sense to start with the
simpler framework of ideal MHD.

\subsection{Equations}

The system of ideal relativistic MHD includes the continuity equation
                                                                                       \begin{equation}
\Pd{t}(\alpha\sqrt{\gamma}\rho u^t)+ \Pd{i}(\alpha\sqrt{\gamma}\rho u^i)=0,
\label{cont1}
\end{equation}
where $\rho$ is the rest mass density of matter and $u^\nu$ is its 4-velocity;
the energy-momentum equations

\begin{equation}
\Pd{t}(\alpha\sqrt{\gamma}T^t_{\ \nu})+ \Pd{i}(\alpha\sqrt{\gamma}T^i_{\ \nu})=
\frac{1}{2} \Pd{\nu}(g_{\alpha\beta}) T^{\alpha\beta} \alpha\sqrt{\gamma},
\label{en-mom1}
\end{equation}
where $T^{\nu\mu}$ is the total stress-energy-momentum tensor; the induction
equation,
\begin{equation}
(1/c)\Pd{t}(B^i)+e^{ijk}\Pd{j}(E_k) =0,
\label{ind1}
\end{equation}
and the divergence free condition

\begin{equation}
  \Pd{i}(\sqrt{\gamma} B^i) =0 .
\end{equation}
The total stress-energy-momentum tensor, $T^{\mu\nu}$, is a sum of the
stress-energy momentum tensor of matter
\begin{equation}
   T_{(m)}^{\mu\nu} = wu^\mu u^\nu -p g^{\mu\nu},
\end{equation}
where $p$ is the thermodynamic pressure and $w$ is the enthalpy per unit volume,
and the stress-energy momentum tensor of the electromagnetic field
\begin{equation}
   T_{(e)}^{\mu\nu} = \frac{1}{4\pi}\left( F^{\mu\gamma} F^\nu_{\ \gamma} -
   \frac{1}{4}(F^{\alpha\beta}F_{\alpha\beta})g^{\mu\nu} \right),
\end{equation}
where $F^{\nu\mu}$ is the Maxwell tensor of the electromagnetic field.  In the
limit of ideal MHD
\begin{equation}
  E_i=e_{ijk}v^jB^k/c,
\label{perf-cond}
\end{equation}
where $v^i=u^i/u^t$ is the usual 3-velocity of plasma.

\begin{figure*}
\fbox{\includegraphics[width=85mm]{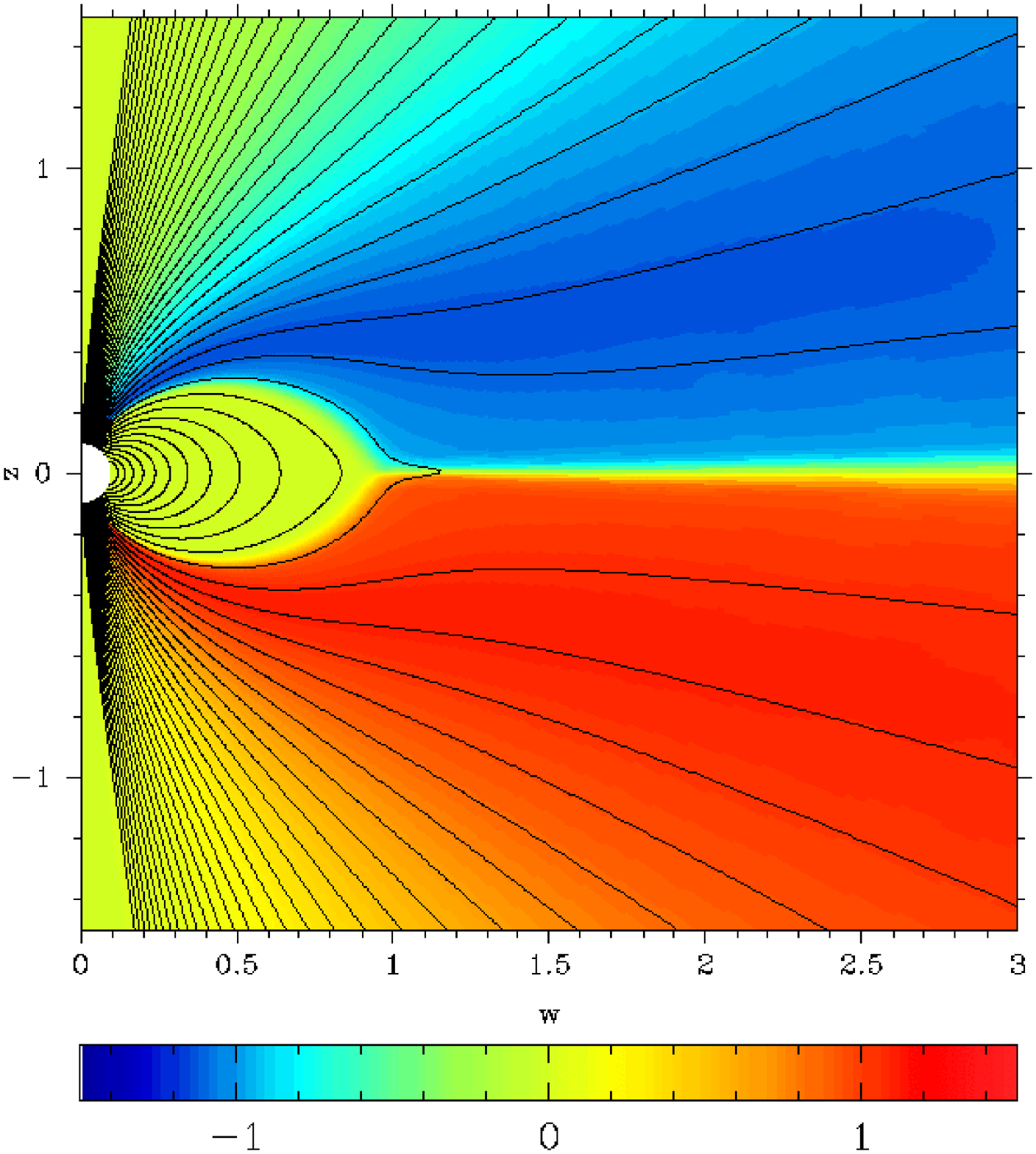}}
\fbox{\includegraphics[width=85mm]{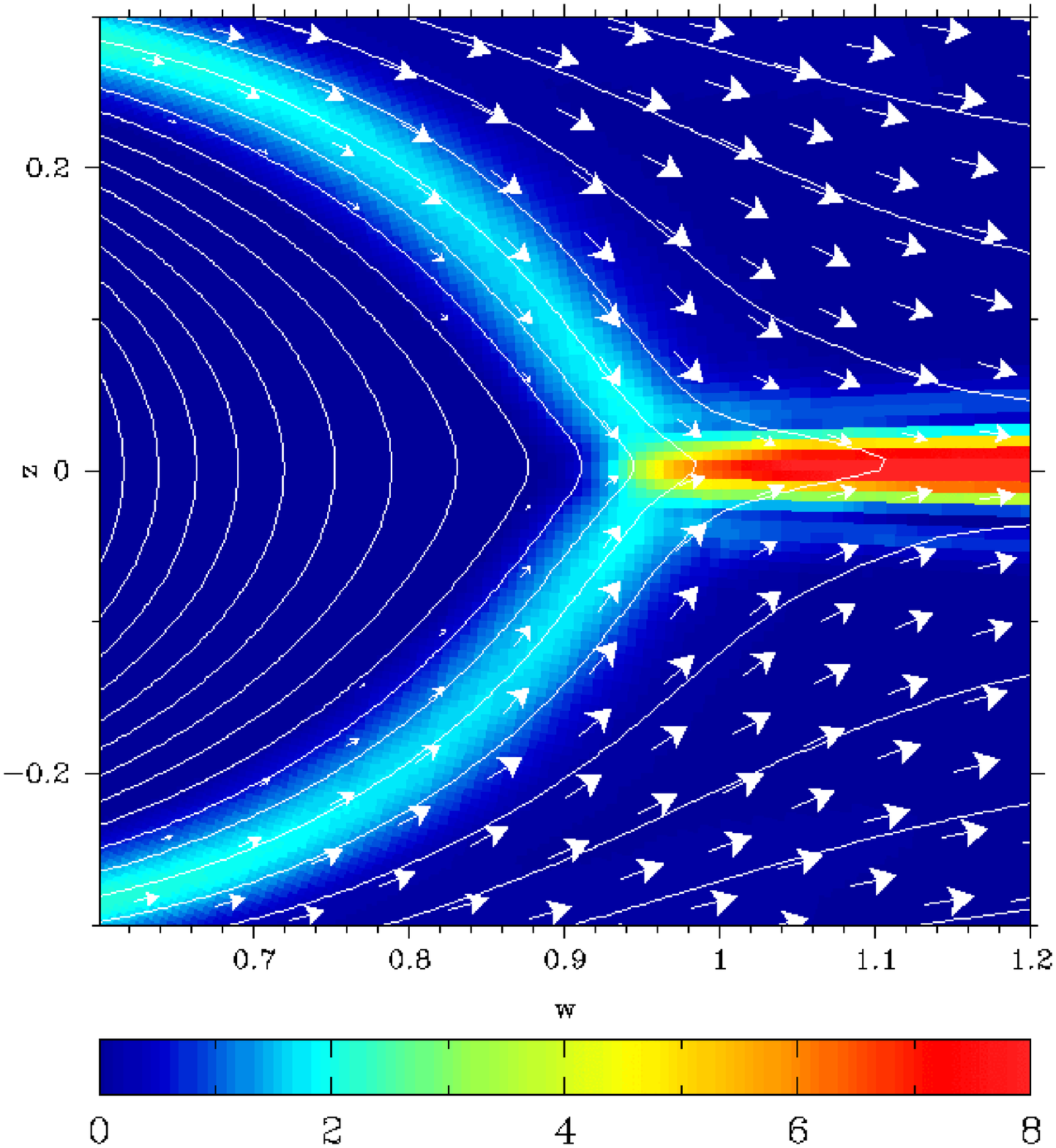}}
\fbox{\includegraphics[width=85mm]{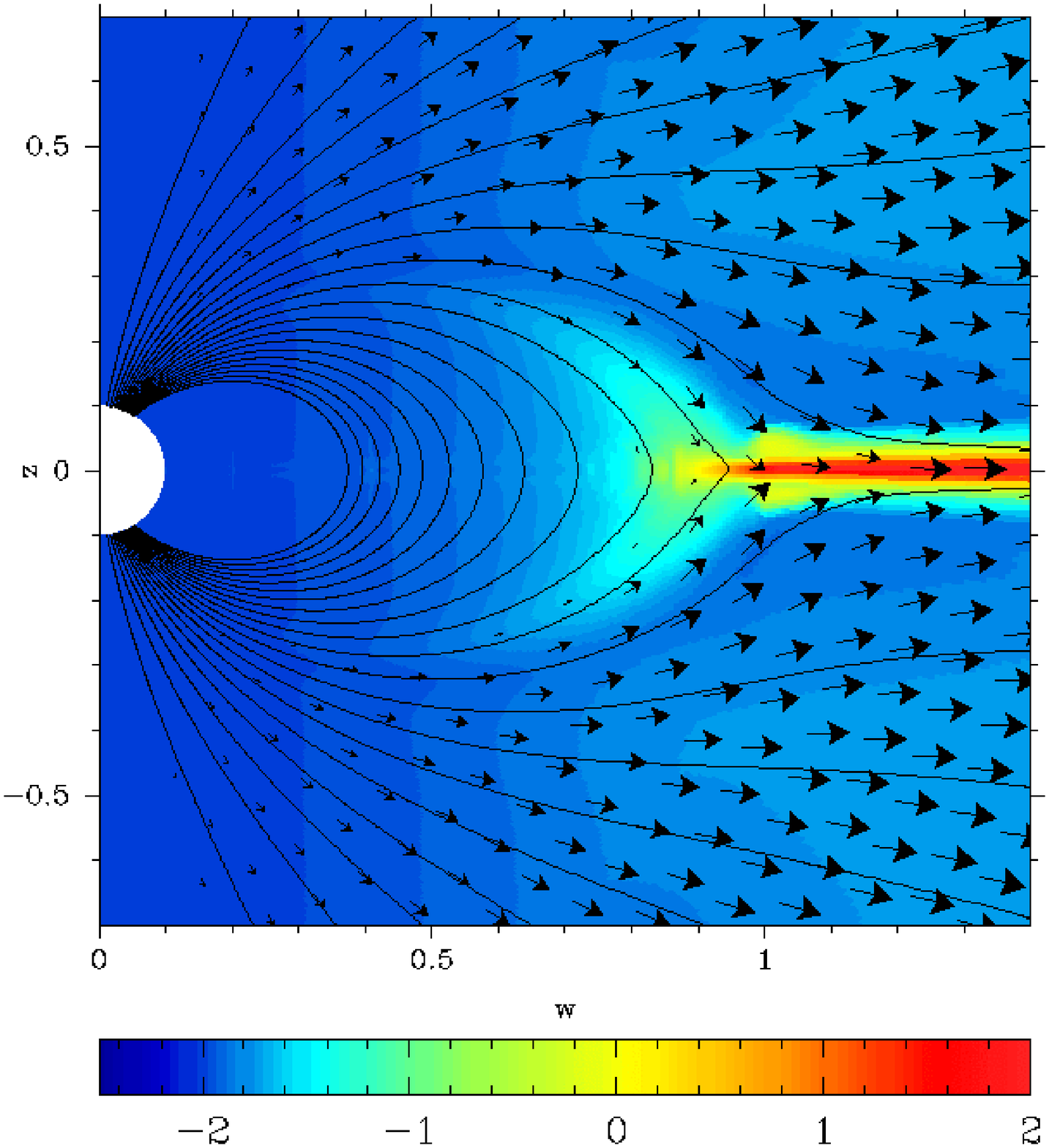}}
\fbox{\includegraphics[width=85mm]{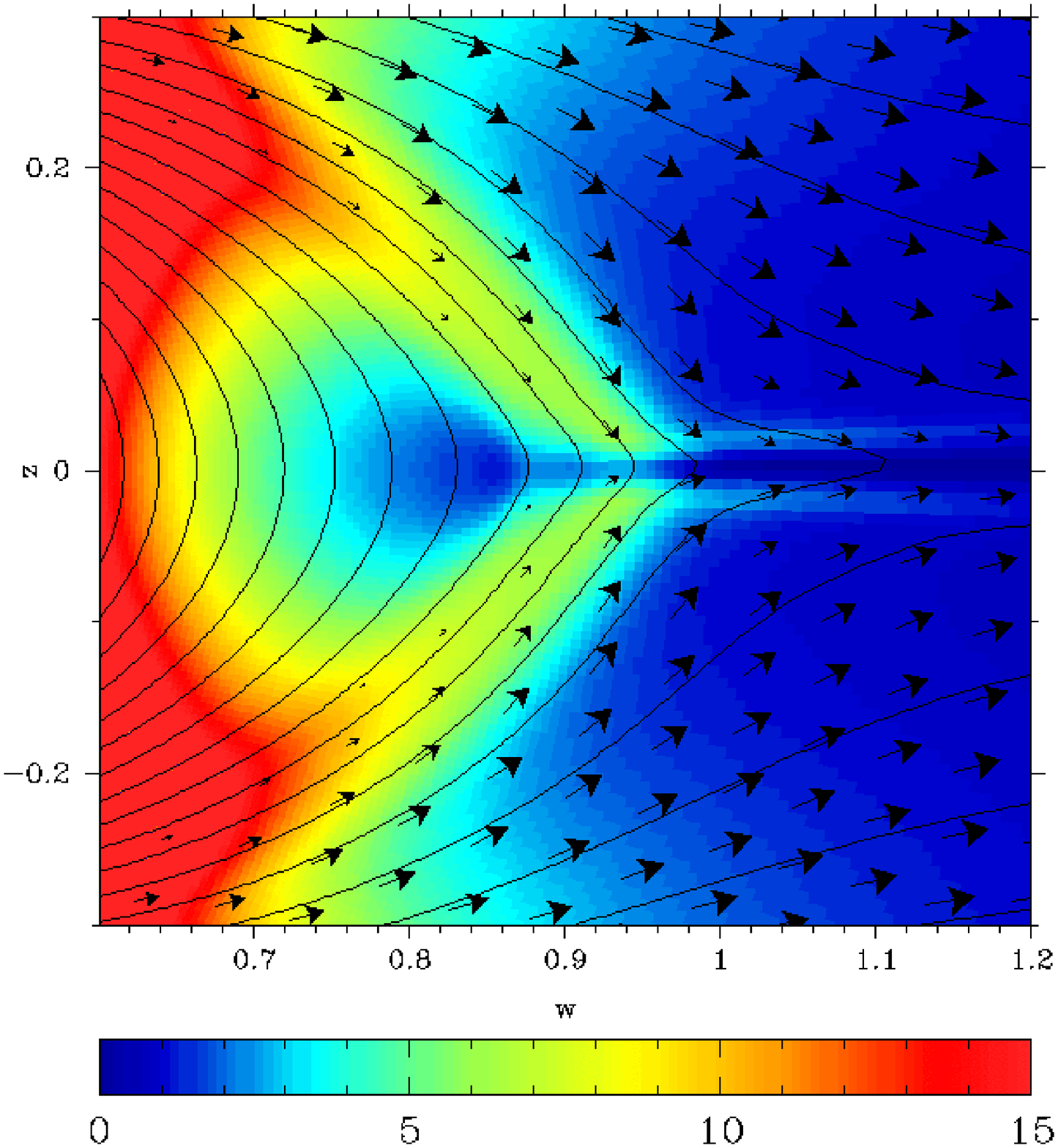}}
\caption{ Inner part of the solution at $t=55$.  
{\it Top left panel:} 
The contours show the magnetic flux function, $\Psi$, and the colour 
image shows $H_\phi$;  
{\it Top right panel:} 
The contours show the magnetic flux function, $\Psi$, the arrows show 
the flow velocity, and the colour image shows the magnitude of the 
poloidal electric current density multiplied by $r^2$; 
{\it Bottom left panel:} 
The contours show the magnetic flux function, the arrows show the flow 
velocity, and the colour image shows the $\log_{10}(wW^2/B^2)$; 
{\it Bottom right panel:} 
The contours show the magnetic flux function, the arrows show the flow 
velocity, and the colour image shows $B^2$.}  
\label{inner}
\end{figure*}

\subsection{Numerical method}

The MHD simulations were carried out using a Godunov-type scheme that is
described in Komissarov\shortcite{Kom99,Kom04b}. However, we had to introduce a
number of additional features in order to overcome a number of challenging
problems specific to the case of highly-magnetized plasma.

The MHD equations become stiff in magnetically dominated domains and this is
exactly the case for the main volume of pulsar magnetospheres where the energy
density of matter is many orders of magnitude less than the one of the
electromagnetic field. There is no hope of reaching such conditions with our
numerical method. However, the electromagnetic part of the MHD solution should
not be very different even when the energy ratio is artificially increased up to
0.1-0.01 -- errors of order of few percent are quite acceptable at this stage of
investigation. In fact, the results of force-free and MHD simulations of the
monopole magnetospheres of black holes strongly support this conclusion
\cite{Kom01,Kom04b}. In those MHD simulations, additional plasma was pumped in
the regions where its energy density had reached a certain lower limit. Here, we
apply a similar trick. The actual condition is
\begin{equation} 
wW^2 \ge a_{(1)}B^2,
\label{trick1} 
\end{equation}
where $W$ is the Lorentz factor of plasma and $a_{(1)}$ is a small constant (we
used $a_{(1)}=0.01$.)  When this condition is broken we increase both $\rho$ and
$p$ in the same proportion so that $wW^2 = a_{(1)}B^2$.

However, the dipolar magnetospheres are more challenging compared to the
monopole ones because of the faster decline of magnetic field strength with the
distance from the star and the existence of dead zones.  Within the dead zones
the magnetospheric plasma is supposed to be in static equilibrium. Thus, the
component of the centrifugal force acting along the magnetic field lines has to
be balanced by some other force.  It has been argued that the dead zone plasma is
charge-separated and the force balance is achieved by means of a small parallel 
component of electric field (e.g. \cite{HoP,Mes}). However the recent eclipse 
observations of the binary pulsar JO737-3039 allowed direct measurement of the particle 
density in the dead zone of one of the components \cite{LyT}. 
It has turned out to be many orders of magnitude higher than the expected density of 
the charged-separated plasma. Whether such a high density is specific for binary systems 
only, as proposed in  \cite{LyT}, or typical for single pulsars as well remains to be 
seen. In any case, the charge-separated dead-zones cannot be modelled within the MHD 
approximation where the required force balance can only be achieved by means of gas 
pressure. This would require the gas pressure 
to increase along the magnetic field line and reach a maximum in the equatorial 
plane. On the contrary, the magnetic pressure of dipolar magnetosphere decreases 
with distance  as $r^{-6}$  and thus the ratio of magnetic to gas pressure has to 
decline even faster than $r^{-6}$. Given the requirement on the dead zone to 
remain magnetically dominated everywhere this implies very high pressure ratio
near the star surface - much higher than our numerical scheme can accommodate.     
To overcome this problem one could consider the case where the star radius is
only 2-3 times smaller than the radius of its light cylinder but this would have 
a rather strong effect on the structure of the magnetosphere.
This is why we have preferred a different strategy. 

Its idea is to
reduce, somewhat arbitrarily, the dynamical effect of the centrifugal force on the
motion of plasma near the star so that it does not destroy the nearly force-free
equilibrium across the magnetic field lines within the dead zone.  One way of
achieving this is to reset the gas pressure and its rest mass density to some
rather low 'target' values, $p_s$ and $\rho_s$, within the dead zone every time
step.  At the same time one may reset the flow speed along the magnetic field
lines to zero, just as it should be in equilibrium. However, the dead zone is
not a well defined region in the case of time-dependent magnetospheres. Instead,
one could apply the same procedure to a volume that is guaranteed to include if
not the whole dead zone then at least its inner part. For example, we used a
sphere of radius $r_s\leq\varpi_{lc}$. This however leads to another
complication -- in the open field region bounded by the sphere there emerges a
strong rarefaction wave, so strong that the code crashes. Fortunately, this can
be avoided if instead of resetting the target values one introduces a relaxation
towards them with the relaxation time gradually increasing towards the boundary
of the relaxation domain. We evolved $p$, $\rho$, and $\bv_\parallel$ according
to the following equations
 
\begin{equation}
   \oder{\bv_\parallel}{t} = -b_{(1)}\bv_\parallel,
\end{equation}
\begin{equation}
   \oder{p}{t} = -b_{(2)}(p-p_s),
\end{equation}
\begin{equation}
   \oder{\rho}{t} =-b_{(2)}(\rho-\rho_s),
\end{equation}
where
\begin{equation}
    b_{(1)}(r)=b_{(0)} f(r),
\end{equation}
\begin{equation}
    b_{(2)}(r,\theta)=b_{(1)} |\cos\theta|,
\end{equation}
\begin{equation}
    f(r) = \left\{
    \begin{array}{rl} 
       (r_{s}-r)/(r_{s}-r_*) & \text{if} r < r_{s} \\ 0 & \text{if} r > r_{s}
    \end{array} \right. , 
\end{equation}
where $b_{(0)}$ is constant, and $r_*$ is the star radius. As one can see, the
relaxation time becomes infinite at $r=r_s$. The results presented below
correspond to $r_s=\varpi_{lc}$. However, we have also tried smaller values of
$r_s$ in order to verify that this does not lead to qualitatively different
results (We discuss the effects of reducing $r_s$ in Section 5.)  The dependence
of $b_{(1)}$ on the azimuthal angle was introduce in order to reduce the
possible adverse effect on the current sheet should it be formed inside the
light cylinder.  The actual value of $b_{(0)}$ is to be found by the method of
trial and error. Finally, we use the following targets for the pressure and
density

\begin{equation}
    p_s = a_{(2)} \rho_s c^2 , \, \rho_s c^2 =a_{(1)} B^2,
\end{equation}
where $a_{(1)}=0.01$ and $a_{(2)}=0.001$.

In these simulations, the computational grid covered the axisymmetric domain
$(r,\theta)=[0.1,50]\times[0,\pi]$ and hence the star radius was set to
$r_*=0.1$. In order to speed up the calculation we started with a relatively low
resolution grid, $124\times61$, and then increased the resolution twice after
the solution seemed to have reached a steady-state on the scale of several
$\varpi_{lc}$. Hence the final grid had $496\times 244$ cells.

\subsection{Results}

The initial solution described a non-rotating magnetosphere with dipolar
magnetic field, eq.(\ref{dipole}), $v^i=0$, $\rho=\rho_s$, and $p=p_s$.  At
$t=0$ the star rotation is switched on and a torsional Alfv\'en wave is emitted
from its surface. As it propagates away larger and larger portion of the
magnetosphere is set into rotation and develops an electric current system. This
process is illustrated in figure \ref{glob} where the contours show the magnetic
field lines and the colour image shows the distribution of $H_\phi=2(I+I_d)/c$,
where I is the total electric current and $I_d$ is the total displacement
current through the circular contour of cylindrical radius $\varpi=r\sin\theta$.
Unity corresponds to $H_\phi = \mu \Omega^2 c^{-2}$.

Behind the wave the solution gradually approaches a steady state.  Figure
\ref{inner} shows the inner region of this steady state solution at $t=55$, the
termination time of the simulations.  The structure of magnetic field lines
suggests that the dead zone extends all the way up to the light cylinder. Beyond
the light cylinder the poloidal magnetic field becomes radial, as expected
\cite{Ing,Mic74}.  Some magnetic field lines are closing up beyond the light
cylinder but they do so within the equatorial current sheet due to finite
artificial resistivity in the numerical scheme -- it is easy to see the
transition between the dead zone and the current sheet within which the magnetic
field lines are highly stretched in the radial direction. The colour image in
the top left panel of this figure shows the distribution of $H_\phi$. Since in a
steady state the displacement current vanishes we have $ H_\phi=2I/c ={\cal A}.$
This image confirms our conclusion on the extension of the dead zone which is
seen in the image as a toroidal structure with $H_\phi=0$ (Indeed, within the
dead zone the poloidal currents do not flow and $H_\phi$ must be zero.) The
jumps in $H_\phi$ occurring at the boundary of the dead zone and along the
equator are indicators of thin sheets of return current.  This image also shows,
though not as clear, that $H_\phi$ and hence $I$ reach maximum amplitude at some
finite distance from the equatorial plane. This indicates the presence of an
additional layer of return current that surrounds the equatorial current sheet
and the dead zone \cite{CKF}.

The current sheets are most prominent in the right panel of figure~\ref{inner}
that shows the distribution the poloidal current density, $\bJ_p$, multiplied by
$r^2$.  One can also see that the thickness of the current sheets near the
Y-point is about $0.5\div0.6\varpi_{lc}$ and that equatorial current sheet
extends inside the light cylinder by approximately the same distance.

\begin{figure}  \hskip 5mm 
\includegraphics[width=75mm,angle=-90]{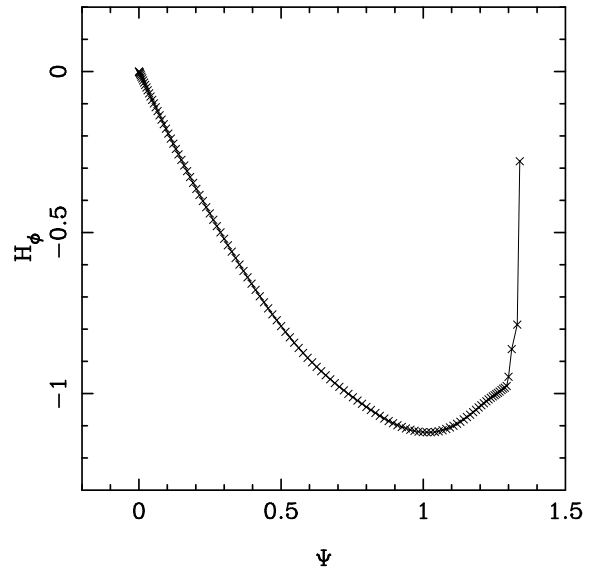}
\caption{$H_\phi$ as a function of the magnetic flux function $\Psi$ on the
sphere $r=1.1$ at time $t=55$.}
\label{current1.2}
\end{figure}

From time to time it was suggested that particle inertia may actually become
dynamically important near the light cylinder, thus rendering the force-free
approximation as unsuitable. The argument develops like this.  Provided the
magnetospheric plasma simply co-rotated with the star its speed would exceed the
speed of light beyond the light cylinder.  This does not occur because of the
rapidly increasing inertial mass of this plasma near the light cylinder. This
leads to very strong centrifugal force that makes the plasma to flow across the
light cylinder thus opening up the magnetic field lines and sweeping them back.
However, this not what occurs in our simulations. Indeed, the top left panel of
figure~\ref{inner} shows that, very much in agreement with the force-free
models, $H_\phi$ is constant along the magnetic flux surfaces even when they
cross the light cylinder. Moreover, this conclusion is fully supported by the
data presented in the bottom left panel of figure~\ref{inner} that shows that
the distribution of $wW^2/B^2$, the quantity that can be used to describe the
relative importance of the inertial effects as well as of the gas pressure.  One
can see that it remains very low everywhere outside of the equatorial current
sheet including the light cylinder.

There is, however, one location in the force-free solution of Contopoulos et
al.\shortcite{CKF} where the inertial effects are expected to become
important. It is the so-called Y-point, that is the point where the dead zone
approaches the light cylinder. Indeed, since the dead zone co-rotates with the
star then plasma particles attached to its field lines rotate with the speed
reaching the speed of light at the y-point. Moreover, one may argue that the
divergence of magnetic field strength at the Y-point discovered by
Gruzinov\shortcite{Gru05} also suggests a breakdown of the force-free
approximation.  This fully agrees with the data presented in the bottom right
panel of figure~\ref{inner} which shows the distribution of $B^2$ in our MHD
solution.  Although the plot shows a local minimum at $(z,\varpi)=(0,0.85)$ and
then some growth of $B^2$ in the direction of the Y-point this growth never
develops.  Moreover,  in our solution the poloidal magnetic
field lines approach the Y-point at an angle of $\simeq 50^o$ to the
equatorial plane (see fig.\ref{inner}), which is significantly lower than 
$77^o.3$ predicted by Gruzinov\shortcite{Gru05} and quite close to the value of 
$56^0.5$ found in Goodwin et al.\shortcite{Goo}. In addition to the inertial effects, 
the relatively high thickness of the current sheets near the Y-point (the right top 
panel of figure~\ref{inner})  and finite gas pressure also contribute to these 
discrepancies between our simulations and the force-free solution. Our results however 
do not prove that the asymptotic force-free solution for the Y-point is irrelevant.  
It seems possible that for sufficiently high magnetization and small resistivity the
exact MHD solution will first approach the force-free asymptote found by Gruzinov and
then deviate from it on smaller scales. This, however, requires further 
investigation.

\begin{figure*}
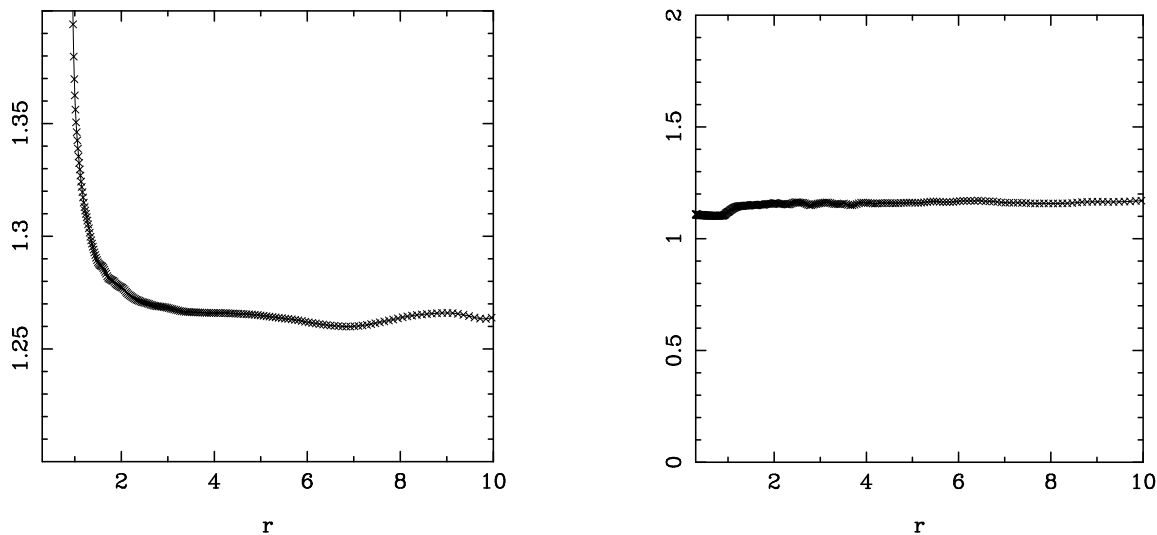

\includegraphics[width=70mm,angle=-90]{figures/mflux.ps}
\hskip 2cm \includegraphics[width=70mm,angle=-90]{figures/eflux.ps}
\caption{ {\it Left panel:} The magnetic flux function, $\Psi$, in the
equatorial plane as a function of $r$ at time $t=55$. {\it Right panel:} The
total energy flux through a sphere of radius $r$ at time $t=55$.  }
\label{fluxes}
\end{figure*}
\begin{figure*}
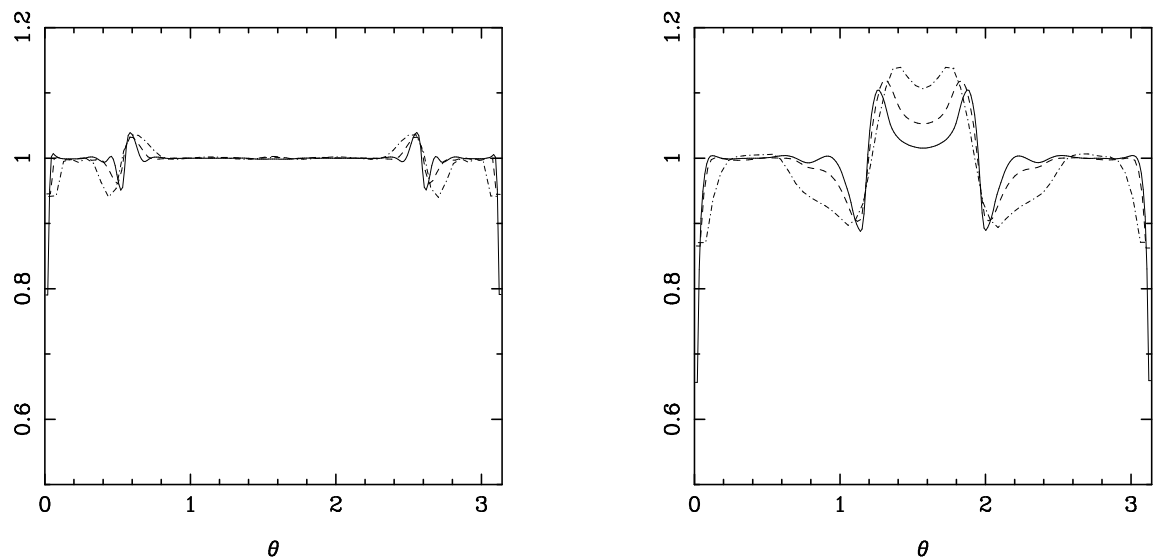

\includegraphics[width=73mm,angle=-90]{figures/om02.ps}
\hskip 2cm \includegraphics[width=73mm,angle=-90]{figures/om07.ps}
\caption{The angular velocity distribution at $t=55$ within the light cylinder
at $r=0.2$ (left panel) and at $r=0.7$ (right panel). The dot-dashed, dashed,
and solid lines show the solutions with 61, 122, and 244 cells in the
$\theta$-direction. }
\label{omega}
\end{figure*}

Figure \ref{current1.2} shows $H_\phi$ as a function of $\Psi$ at $r=1.1$ and
allows us to see these details of the current system more clearly.  This
distribution is very similar to the one given in figure 4 of Contopoulos et
al. (1999). The minimum has
$$ \Psi_{min} \simeq 1.02\frac{\mu\Omega}{c} \text{and} A_{min} \simeq
1.12\frac{2\mu\Omega^2}{c^2}.
$$

The left panel of figure \ref{fluxes} shows the distribution of magnetic flux
function $\Psi$ in the equatorial plane of the final solution.  As a consequence
of the finite artificial resistivity in our numerical scheme the magnetic field
lines continue to close up even beyond the line cylinder and this shows itself
via a systematic decline of $\Psi$ at $r>1$. Thus, it is not so straightforward
to determine the fraction of opened field lines in this solution. One way to
describe it quantitatively is by giving the value of the flux function exactly
at $(\theta=\pi/2,r=1)$. This gives us
\begin{equation} 
   \Psi_{yp} \simeq 1.37 \div 1.38 \frac{\mu\Omega}{c}.
\end{equation} 
On the other hand, figure~\ref{fluxes} shows that the decline of $\Phi$
significantly slows down at $r>2$ where $\Phi$ reaches the value of
\begin{equation} 
     \Psi_{open} \simeq 1.26 \div 1.27 \frac{\mu\Omega}{c}.
\end{equation}
These numbers should be compared with the $c\Psi_{open}/\mu\Omega = 1.23 $ in
Contopoulos\shortcite{Con05} and Timokhin\shortcite{Tim}, $1.27$ in
Gruzinov\shortcite{Gru05}, and $1.36$ in Contopoulos et al.\shortcite{CKF}.

The right panel of figure \ref{fluxes} shows the total flux of energy through
the sphere of radius $r$ that should be constant in a steady state solution. One
can see that it is indeed more or less constant with the exception of $1<r<2$
where the energy flux slightly increases.  This increase is a permanent feature
that arises because our scheme is not strictly conservative.  In order to
evaluate the spin-down power of the star we used the energy flux through the
sphere of radius $r=1$, thus ignoring the non-physical increase of total
luminosity beyond $r=1$. This gives us
\begin{equation} 
     L \simeq 1.1 \frac{\mu^2\Omega^4}{c^3}
\end{equation}
which is in a very good agreement with the result by Gruzinov(2005).

Figure \ref{omega} shows the distribution of the angular velocity of magnetic
field lines, $\Omega_f$.  In the exact steady-state force-free solution the
magnetic field lines rotate with same angular velocity as the star,
\begin{equation} 
\Omega_f=\Omega.
\end{equation} 
However in our solution, $\Omega_f$ noticeably deviates from $\Omega$ inside the
layer coincident with the current sheet between the open field lines and the
dead zone (see fig.\ref{omega}). The most likely reason for this is the enhanced
numerical resistivity in the current sheet.  The comparison of solutions with
different numerical resolution (fig.\ref{omega}) shows that the thickness of the
layer significantly decreases with resolution.  However, the amplitude of the
perturbations does not seem to depend on the resolution.

\begin{figure}
\fbox{\includegraphics[width=81mm]{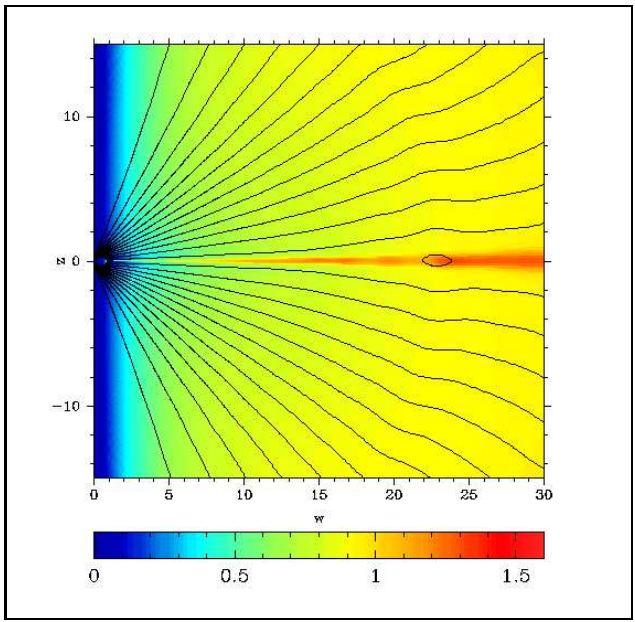}}
\caption{The wind zone structure of a dipolar magnetosphere. The contours show
the field lines of poloidal magnetic field. The colour image shows the
distribution of $\log_{10} W$}
\label{monop}
\end{figure}

\begin{figure*}
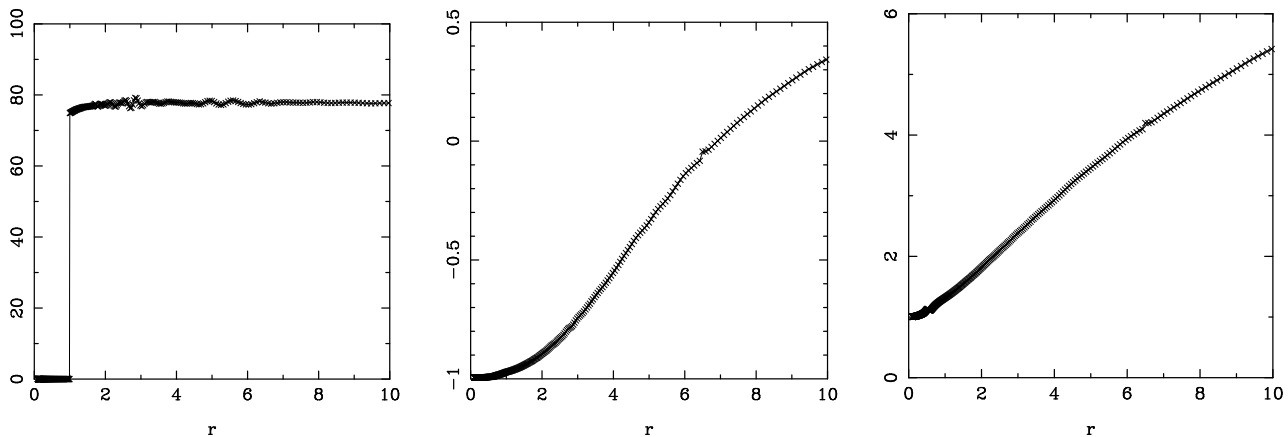

\includegraphics[width=57mm,angle=-90]{figures/sigma.ps}
\hskip 5mm \includegraphics[width=57mm,angle=-90]{figures/fast.ps}
\hskip 5mm \includegraphics[width=57mm,angle=-90]{figures/lor.ps}
\caption{ The variation of a number key parameters along the ray $\theta=1$.
{\it Left panel:} The local magnetization parameter $\tilde{\sigma}$; {\it Middle
panel:} The wavespeed of the ingoing fast wave in the radial direction (the
speed of light c=1); {\it Right panel:} The Lorentz factor of the wind.  }
\label{fast}
\end{figure*}

Figure~\ref{monop} shows that far away from the star the distribution of
poloidal field lines is very similar to the split-monopole one. This is exactly
what was concluded in the pioneering papers by Ingraham\shortcite{Ing} and
Michel\shortcite{Mic74} on the force-free magnetospheres. More recently, the
centrifugally driven outflows in split-monopole magnetospheres have been studied
within the cold-MHD approximation.  According to these studies the Lorentz
factor of centrifugally accelerated plasma at the fast critical point is
\begin{equation} 
  W = \sigma^{1/3},
\label{beskin}
\end{equation}  
where $\sigma$ is the so-called magnetization parameter which is defined as the ratio of
the Poynting flux density to the rest mass-energy flux density at the foot point of 
the magnetic field line \cite{Bes97}. For outflows that are initially Poynting 
dominated $\sigma \simeq \tilde{\sigma}$, where $\tilde{\sigma}$ is the ratio of the 
total energy flux density to the rest mass-energy flux density. In contrast to $\sigma$, 
$\tilde{\sigma}$ can be measured at every point of the field line as it is constant 
along it.   Figure~\ref{fast} shows the distribution of $\tilde{\sigma}$,
$W$, and the ingoing speed of the fast wave in the radial direction along the
ray $\theta = 1$rad. From these data we find that along this ray $\sigma \simeq
77$ and the fast point is located at $r\simeq 6.8$ where $W \simeq 4.3$. On the
other hand, we can use the above value of $\sigma$ in order to calculate the
Lorentz factor at the fast point directly from eq.~\ref{beskin} -- this give us
$W=4.25$. Thus, the split-monopole model provides quite a good model for the
wind zone of aligned magnetic dipole.

Another interesting feature of figure~\ref{monop} is the significantly higher
value of the Lorentz factor of the outflow within the equatorial current
sheet. This suggests that the mechanism of the flow acceleration is somewhat
different there. The obvious suspect is heating due to resistive dissipation of
the electromagnetic energy. The high value of Lorentz factor in the current
sheet suggests that its contribution into the global energy transfer can be
quite significant.

\begin{figure*}
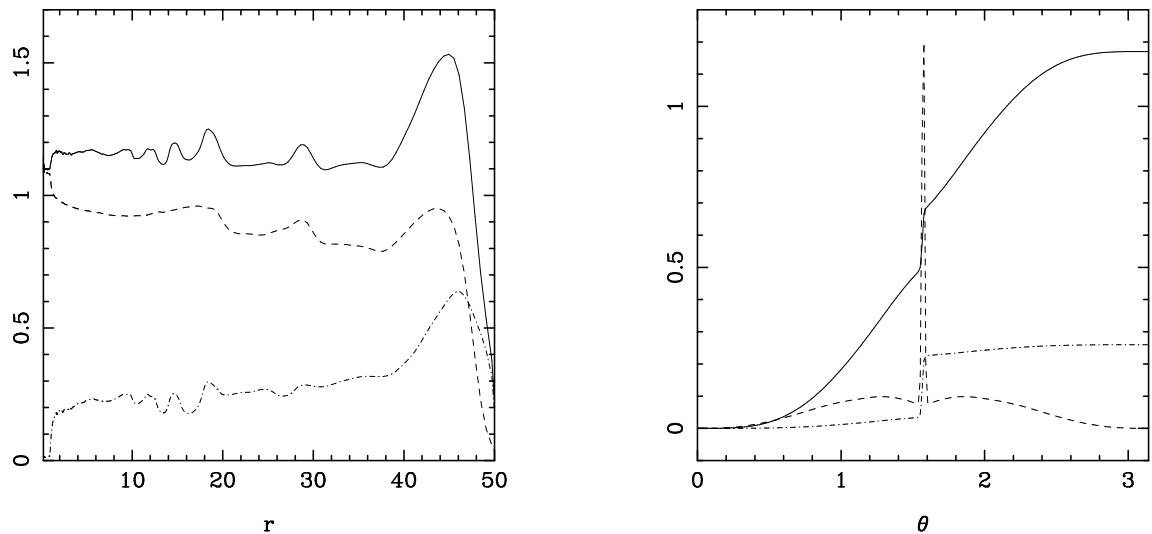

\includegraphics[width=70mm,angle=-90]{figures/convert.ps}
\hskip 2cm \includegraphics[width=70mm,angle=-90]{figures/spike.ps}
\caption{Conversion of the electromagnetic luminosity into the hydrodynamic
luminosity of the wind.  {\it Left panel:} The total luminosity (solid line),
the electromagnetic luminosity (dashed line), and the hydrodynamic luminosity
(dot-dashed) of the wind at $t=55$.  {\it Right panel:} The angular distribution
of the total luminosity at $r=10$. The solid line and the dot-dashed line show
the total luminosity and the hydrodynamic luminosity within the polar cone of
angle $\theta$ respectively; the dashed line shows the total flux density in the
radial direction.  }
\label{convert}
\end{figure*}

The left panel of figure~\ref{convert} shows the evolution of the wind
luminosities with the distance from the star by the end of simulations. The
total luminosity is more or less constant apart from noticeable perturbations
around $r=17$, $r=30$, and $r=45$.  The big bump around $r=45$ is related to the
leading front of the wind. The perturbations around $r=17$ and $r=30$ are
related to the grid refinement events -- each time the computational grid is
refined the numerical solution evolves to a slightly different state, most of
all in the inner region of the computational domain.  This triggers noticeable
waves propagating away from the star.  The electromagnetic luminosity, which is
shown in figure~\ref{convert} by the dashed line, gradually decreases with
distance thus indicating the ongoing conversion of the electromagnetic energy
into the hydrodynamic energy. This is supported by the evolution of the
hydrodynamic luminosity of the wind which is shown by the dash-dotted line. One
of the interesting feature of this line is its rapid rise initial rise that
suggests a particularly effective energy conversion. The nature of this
conversion is clarified in the right panel of figure~\ref{convert} that shows
that the current sheet accounts for about 70\% of the total hydrodynamic
luminosity at $r=10$ and this corresponds to about 15\% of the total wind
luminosity.  This clearly points out at the Ohmic heating in the current sheet
as the main source of the energy conversion.

\section{Discussion} 

There is not much to discuss in connection with our force-free simulations of
pulsar magnetospheres. We simply have not been able to make the required
progress using this framework because of the inability to handle the equatorial
current sheet. For this reason we will focus in this section almost entirely on
the results of our MHD simulations and their possible implication for pulsar
physics.
  
One of the key goals of this study was to determine there the MHD equations
allow stable, or quasi-stable, steady-state solutions for dipolar axisymmetric
magnetospheres of neutron stars. This problem had become particularly
interesting since the discovery a whole family of steady-state force-free
solutions continuously parametrised by the location of their y-point
\cite{Goo,Tim}.  Our results indicate the existence of a unique steady-state MHD
solution to the problem and this solution is very close to the force-free
stationary solution of the pulsar equation with the y-point located at the light
cylinder -- the original solution of pulsar found by Contopoulos et
al.\shortcite{CKF}. Why this solution is preferable to those whose y-point is
located well inside the light cylinder has already been explained in
\cite{Con05}.  If resistivity is not vanishingly small then even if the initial
solution had the dead zone buried well inside the light cylinder it would take
only a finite time for the anti-parallel field lines of the current sheet
between the y-point and the light cylinder to reconnect and to form closed loops
that become a part of the dead zone.  The reconnection should also occur beyond
the light cylinder but there the outflow is super-Alfv\'enic so the net outcome
of the reconnection is likely to be the development of magnetic islands carried
by the wind away from the star \cite{Uzd04}.  The rate of the reconnection depends
on the actual resistivity in the current sheet, and in our simulations the
resistivity is purely artificial.  However, the ultimate outcome is unlikely to
be different. Indeed, since the particle inertia on the closed field lines
located well inside the light cylinder of pulsar magnetospheres is extremely
small there is no restoring force that would make this field lines to open up
again. This is supported by the fact that the total electromagnetic energy 
of the force-free magnetosphere is minimum when the Y-point is located on the light 
cylinder \cite{Tim}.   
 
\begin{figure}
\fbox{\includegraphics[width=81mm]{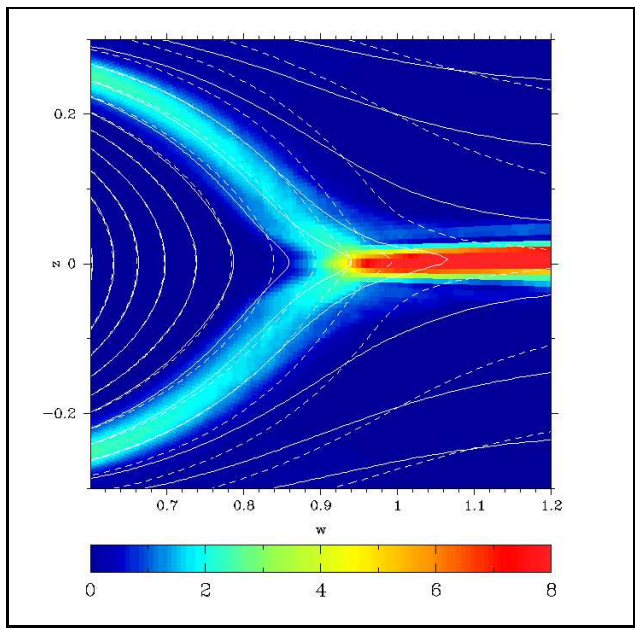}}
\caption{Comparison of solutions with relaxation domains of 
different sizes. The colour image shows $H_\phi$ and the solid lines
show the magnetic flux function for the model with $r_s=0.7\varpi_{lc}$. 
The dashed lines show the magnetic flux function for the model 
with $r_s=\varpi_{l}$.} 
\label{forcing}
\end{figure}

At this point it makes sense to discuss whether the relaxation procedure applied
within $r_s=\varpi_{ls}$ (see Sec.4.2) could somehow promote the expansion of
the dead zone towards the light cylinder. As we have already pointed out in
Sec.4.2 all relaxation times gradually increase towards infinity as
$r\rightarrow r_s$. Thus, near the light cylinder the effect of the relaxation
procedure on the flow is increasingly small. Moreover, the relaxation time for
the gas pressure becomes infinite at $\theta = \pi/2$.  This would allow the
build up of gas pressure required to support the equatorial current sheet should
it exist within the light cylinder.  However, in order to fully resolve this
issue we carried out additional simulations with the radius of the relaxation
sphere pushed down to $r_s=0.7\varpi_{lc}$. Figure~\ref{forcing} allows to
compare both solutions with regard to the location of the Y-point. The colour
image in this figure shows the distribution of $H_\phi$ and the solid lines show
the magnetic surfaces for the model with $r_s=0.7\varpi_{lc}$ whereas the dashed
lines show the magnetic surfaces for the model with $r_s=\varpi_{lc}$. As one
can see more field lines open up for smaller $r_s$ and this is caused by the
increased dynamical role of the centrifugal force.  However, the location of
Y-point remains basically the same. Thus, our relaxation procedure cannot be
considered as the reason for the dead zone extending all the way up to the light
cylinder.

Contopoulos\shortcite{Con05} also pointed out that the open field lines of
pulsar magnetospheres may rotate at a slower rate than the closed field lines of
the dead zone due to the finite potential gap of the polar cap. In particular,
he speculated about the possibility of a significant growth of the polar gap due
to a sudden decrease of the particle injection rate in the gap in order to
explain the explosive phenomena like the December 27, 2004 burst in SGR 1806-20.
Although, it is not clear what could cause such a sudden incident of ``charge
starvation'' it was suggested long ago that a slow systematic evolution of the
polar gap could result from the gradual spin-down of the star \cite{Stu,RuS}.
Thus, differentially rotating magnetospheres are not just interesting
theoretical models but can be very much relevant for the real pulsars.

Contopoulos\shortcite{Con05} found force-free stationary numerical solutions for
such magnetospheres in the simple case of a uniformly rotating polar cap and 
a uniformly rotating dead zone which was assumed to corotate with the star.  
(In fact, the dead zone may also have a potential gap that separates electric 
charges of opposite sign but it is expected to be rather small \cite{HoP}.) 
These solutions have two light
cylinders -- a smaller one for the dead zone and and a larger one for the open
field lines and the dead zone extends all the way up to its light cylinder.
Contopoulos\shortcite{Con05} argued that although there existed solutions with a
smaller dead zone they were not sustainable due to reconnection in the part of
the equatorial current sheet that runs between the y-point and the light
cylinder of the dead zone.  

In fact, the numerical models constructed by Contopoulos\shortcite{Con05}
are likely to be globally unstable to reconnection too. Indeed, in these models
the equatorial current sheet continues inside the light cylinder of open magnetosphere 
and magnetic reconnection occurring in this region should lead to creation  
of new closed field lines. However, this case is somewhat more involved. Let us imagine that
such reconnection has indeed occurred.  The field lines that have just closed
down are now beginning to spin-up and corotate with the dead zone. However, they
extend beyond the light cylinder of the dead zone and for this reason they can
not corotate with it -- as they spin-up they begin opening up again. Once have been
opened up they begin to slow down, thus creating conditions for the next closing down event,
and so on. This simple analysis suggests that such differentially rotating
magnetospheres cannot be stationary and have to develop oscillations. The
typical time scale of such magnetospheric oscillations seems to be determined by
the spin-up time that must be comparable with the time required for an Alfv\'en
wave to cross the distance between the light cylinder of the dead zone and the
star back and forward.  Since the Alfv\'en speed is relativistic and the
magnetic field has a significant radial component the crossing time has to be
comparable with with rotational period of the star. The reconnection rate is
more likely to effect the amplitude of these oscillations rather than their
time-scale, a quicker reconnection leading to a larger fraction of magnetic
field lines involved in this process of closing down and opening up. We wont to
speculate that these oscillations may be relevant to the origin of the
sub-pulses of radio pulsars, e.g. \cite{MaT}.  For periodic magnetospheric
oscillations we would have a phenomena reminiscent of beating waves -- this may
explain the so-called drifting sub-pulses. However, one may also expect
quasi-periodic and even chaotic oscillations and they would results in a much
more complicated behaviour of sub-pulses.
     
Our MHD solution has a number of interesting features that could not possibly be
found in the ideal force-free solution of Contopoulos et al.(1999). Some of them
do not depend much on the details of resistivity like, for example, the
centrifugal acceleration of the wind outside of the current sheet. Others, like
the Ohmic dissipation and the wind acceleration in the equatorial current sheet,
do and we have to exercise a reasonable degree of caution when interpreting them
-- MHD simulations with only artificial resistivity can provide at most
qualitatively correct description of such features.  However, it is interesting
that the dissipation of electromagnetic energy in the equatorial current sheet
has already been considered as a promising explanation of the high luminosity
gamma-ray emission from young pulsars \cite{Lyu96,Kir}.  Given the fact that
these gamma-rays carry away a significant fraction of the spin-down power, up to
10\% in the most extreme examples \cite{Tho}, and that pulsar magnetospheres are
highly magnetically dominated, an efficient dissipation of Poynting flux
somewhere in the magnetosphere is needed to explain these observations, and the
equatorial current sheet is one of the most natural locations for such
dissipation.  In particular, Lyubarskii (1996) argued that this high energy
emission originates from the current sheet just beyond the dead zone (This is
exactly where our simulations show most effective Ohmic dissipation.)  The
polarizational observations of the optical emission from the Crab pulsar support
this idea. This emission is found to be polarised parallel to the rotation axis
at the peaks of the pulses and between the pulses (within the pulses the
polarisation vector rotates, which may be attributed to the rotation of the
magnetosphere). This shows that the magnetic field of the emitting region is
predominantly perpendicular to the rotation axis as it should be if this
radiation is generated in the equatorial current sheet.

\section*{Acknowledgements}
I am particularly grateful to Anatoly Spitkovsky and Yuri Lyubarsky for
countless discussions of the various aspects of this problem and appreciate 
the interesting comments made by Leon Mestel, Vasily Beskin, and Maxim Lyutikov 
on the first version of this paper. Finally, this is the right place to thank  
the organises of the KITP program on the {\it Physics of Astrophysical
Outflows and Accretion Discs} (C.Gammie, O.Blaes, H.Spruit) as well as the 
staff of KITP UCSB for the excellent opportunity to work on this problem during 
the spring of 2005 and to exchange views with many experts in the field, including 
Jon Arons, Sergei Bogovalov, Dmitri Uzdensky, and Niccolo Buccantini, from whom 
I have learned a lot.  This research was funded by PPARC under the rolling grant 
``Theoretical Astrophysics in Leeds'' and by the National Science Foundation 
under Grant No. PHY99-07949.

\end{document}